# GaAs/AlGaAs Nanowire Photodetector

Xing Dai,[1,†] Sen Zhang,[1,2,3,†] Zilong Wang,[1,4] Giorgio Adamo,[4] Hai Liu,[5] Yizhong Huang,[5] Christophe Couteau,[2,4,6] and Cesare Soci[1,2,4,*]

[1] *Division of Physics and Applied Physics, School of Physical and Mathematical Sciences, Nanyang Technological University, Singapore 637371*

[2]*CINTRA CNRS-NTU-Thales, UMI 3288, Singapore 637553*

[3]*Tang Optoelectronics Equipment Co., China 201203*

[4]*Centre for Disruptive Photonic Technologies, Nanyang Technological University, Singapore 637371*

[5] *School of Material Science and Engineering, Nanyang Technological University, Singapore 639798*

[6]*Laboratory for Nanotechnology, Instrumentation and Optics, University of Technology of Troyes, 10000 Troyes, France*

† These authors equally contributed to the work.

* Corresponding author: csoci@ntu.edu.sg

**Abstract**

We demonstrate an efficient core-shell GaAs/AlGaAs nanowire photodetector operating at room temperature. The design of this nanoscale detector is based on a type-I heterostructure combined with a metal-semiconductor-metal (MSM) radial architecture, in which built-in electric fields at the semiconductor heterointerface and at the metal/semiconductor Schottky contact promote photogenerated charge separation, enhancing photosensitivity. The spectral photoconductive response shows that the nanowire supports resonant optical modes in the near-infrared region, which lead to large photocurrent density in agreement with the predictions of electromagnetic and transport computational models. The single nanowire photodetector shows remarkable peak photoresponsivity of 0.57 A/W, comparable to large-area planar GaAs photodetectors on the market, and a high detectivity of $7.2 \times 10^{10}$ cm$\sqrt{\text{Hz}}$/W at λ=855 nm. This is promising for the design of a new generation of highly sensitive single nanowire photodetectors by controlling optical mode confinement, bandgap, density of states, and electrode engineering.

Keywords: III-V nanowires, nanowire photodetector, transport in nanowire heterostructures, optical confinement.

Thanks to their quasi-one dimensional geometry and inherently large surface-to-volume ratio, semiconductor nanowires are expected to enhance light confinement and photosensitivity in a variety of optoelectronic devices such as photodetectors,[1-2] solar cells,[3-5] optical switches[6-7] and interconnects.[8-9] Among the nanowire materials, III-V compounds offer clear advantages in bandgap engineering and a high degree of control in bottom-up synthesis and heterostructure formation. Together with their optimal optoelectronic characteristics, notably direct bandgap absorption and high carrier mobility, this creates opportunities to realize III-V nanowire photodetectors with controllable wavelength sensitivity, high response speed, and efficient light-to-current conversion.[1, 10-12] Compared to indirect bandgap material (e.g. Si), direct bandgap III-V materials such as GaAs absorb light more efficiently, thus may achieve equivalent photosensitivity as Si within much smaller volumes, leading to smaller devices. Furthermore, GaAs nanowires can be grown at high growth rates on various substrates, facilitating direct hybrid integration of detectors and driving electronics in telecommunication chips or imaging arrays.[13-15]

One of the main issues with III-V nanowires is that the large density of surface states inherent of their geometry tends to degrade device characteristics by pinning the surface Fermi energy, severely limiting performance of bare nanowire photoconductors and light-emitting devices at room temperature. Surface states act as non-radiative carrier traps and increase the surface scattering, leading to a carrier mobility decrease and potential fluctuation,[16-19] an issue that may be overcome by conformal coating of the nanowires with a "shell" that passivates the core surface and decreases non-radiative carrier traps. This was shown to result in faster photoresponse in GaAs/AlGaAs core-shell nanostructure with shell-to-shell electrodes compared to bare GaAs nanowire.[2] Very recently, with the advantage of surface passivation and spatial carrier confinement at the heterointerface brought by the larger bandgap AlGaAs shell, significant progress has been made on near-infrared single GaAs/AlGaAs nanowire laser working at room temperature.[20] High efficiency solar cells have been achieved in nanopillar-array based on GaAs radial p-n junctions with the InGaP-passivating shell.[21] Superior performance of single nanowires for photovoltaic applications have also been observed in GaAsP p-i-n[3] and GaAs p-i-n[4] radial nanowires with core-to-shell contacts. These structures, however, are difficult to obtain since they require precise control over the doping of the active materials. Another major challenge in nanowire detector technology is posed by the physical contact between the nanowire and the electrode. Indeed, formation of ohmic contacts on some III-V nanowires remains a challenge due to the aforementioned surface Fermi level pinning. Instead, a metal-semiconductor-metal (MSM) configuration, consisting of a back-to-back Schottky diode structure, has been proposed and employed in the fabrication of photodetector devices since 1990s.[22-25] In the dark, the current transport of MSM photodetector is primarily determined and limited by thermionic emission.[26] Under illumination, the increase of carrier density enhances the tunneling probability across the Schottky barrier at the metal/semiconductor interface, where the self-built potential plays a crucial role in the carrier separation and transport, particularly under reverse-bias.[25] As a result, high speed and high sensitivity detection can be achieved.[1-2, 24-25]

In this work we demonstrate a III-V nanowire photodetector structure that combines a radial shell for surface passivation and carrier confinement with a back-to-back Schottky diode structure to avoid non-ohmic contacts. By selectively contacting the core and the shell of the individual nanowire, we achieve MSM configuration with a type I GaAs/high-temperature GaAs/AlGaAs core-multishell nanowire, demonstrating light detection performance comparable to large-area commercial GaAs photodectors at room temperature. Combined analysis of structural, optical and optoelectronic properties is carried out toward a comprehensive understanding of light absorption, mode confinement, and photocurrent transport in these structures.

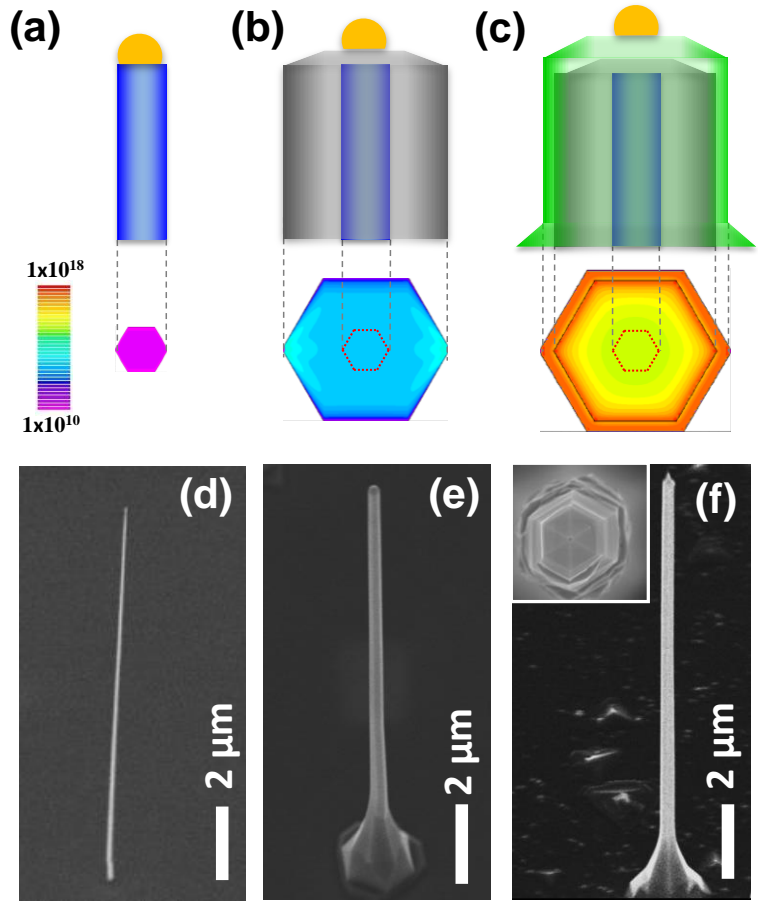

**Figure 1.** Schematic of (a) core GaAs nanowire (r=40 nm), (b) core-shell GaAs/high-T GaAs nanowire (40 nm/170 nm), and (c) core-multishell GaAs/high-T GaAs/AlGaAs nanowire (40 nm/170 nm/30 nm). The corresponding contour plots show the simulated spatial distribution of electron concentration in the cross-sectional plane. Scale bar is in log scale and unit is $cm^{-3}$. (d)-(f) SEM images of nanowires at three different growth stages (obtained with a tilt angle of 45°). The Au nanoparticles used to seed the VLS growth is visible on top of the wires. The inset of (f) is a top-view image of the core-multishell nanowire showing its hexagonal cross-section.

A three-stage core-multishell nanowire growth is implemented to increase thickness of the core, which simulations predict to favor formation of a homogeneous two-dimensional electron "tube" (2DET) at the heterointerface between the GaAs core and the AlGaAs shell, and to create a type-I radial heterostructure (Fig. 1). First, vertical GaAs nanowires with radius of 40 nm were grown by conventional vapor-liquid-solid (VLS) metal-organic chemical vapor deposition (MOCVD) seeded by Au nanoparticles (Figs. 1a and 1d). A thick GaAs overcoating was then deposited at higher growth temperature (Figs. 1b and 1e); the high-temperature (high-T) GaAs layer increases the nanowire core diameter without introducing substantial crystallographic defects, which are typically present in nanowires grown from large diameter nanoparticles (~200 nm).[27] Finally, a relatively thick (to prevent complete Al oxidation) AlGaAs shell was grown to complete the structure (Figs. 1c and 1f). The resulting GaAs/AlGaAs multishell overlayers have thickness of 170 and 30 nm, respectively. The complete core-multishell nanowire structure has total radius of ~240 nm and hexagonal cross-section, as seen in the top-view SEM image shown in the inset of Fig. 1f.[28]

Self-consistent Schrödinger and Poisson equations were solved to map the electron distribution within the hexagonal cross-sectional plane of the bare nanowire and the coaxial structures, and the corresponding contour plots are shown in Figs. 1a-c. Electron contour plots in Fig. 1a and 1b exhibit extremely low electron concentration in undoped GaAs nanowires with the presence of surface states, while with the n-type AlGaAs (doping of $5\times10^{17}$ cm$^{-3}$) high electron density is accumulated at the AlGaAs/GaAs heterointerface, as shown in Fig.1c. Because of the high surface-to-volume ratio, the impact of surface states is pronounced in nanowires and pins the surface Fermi energy within the forbidden band for most III-V compounds (such as GaAs,[29] InP,[29] and GaN[30]). Consequently, a space charge depletion layer exists inside the nanowire. Nanowires with small diameter (Fig. 1a) can be fully depleted, while those with larger diameter (Fig. 1b) can provide a substantial band bending at the surface and preserve a conducting channel.[31-32] By growing an AlGaAs shell over the GaAs core, electrons form a narrower and more uniform distribution at the interface, similar to a two-dimensional electron gas (2DEG) in planar GaAs/AlGaAs heterojunctions, which confines carrier transport in axial direction and eliminates diffusion in the radial path. The high-density, cylindrical 2DET formed along the nanowire axis is expected to promote photoinduced charge dissociation and prolong carrier lifetime, thus increasing photoconductive response of the detector.

The crystalline properties of our core-multishell GaAs/high-T GaAs/AlGaAs nanowire structure were investigated by high-resolution transmission electron microscopy (HR-TEM). HR-TEM images were acquired in a JEOL 2100F microscope using electron acceleration voltage of 200 kV. To obtain cross-sectional structural information, the nanowire was first thinned on opposite sides along the hexagonal edges using focused ion beam (FIB), then cut at its base and transferred onto a TEM copper grid. A representative bright-field TEM image is shown in Fig. 2a, where the boundaries between the three different GaAs, high-T GaAs and AlGaAs shell regions are marked with yellow dashed lines. Corresponding HR-TEM images of core GaAs, high-T GaAs and AlGaAs shell are presented in Fig. 2b-d, with fast Fourier transform (FFT) patterns of each region shown in the insets. All three regions are characterized as a distinct zinc blende (ZB) structure with excellent crystallographic quality. FFT patterns obtained from this sample show sharp spots characteristic of uniform diffraction from [101] zone axis of ZB GaAs and AlGaAs crystals, proving the effectiveness of high temperature overgrowth to increase the thickness of the GaAs core without degrading single crystal quality of the original small-diameter nanowire.

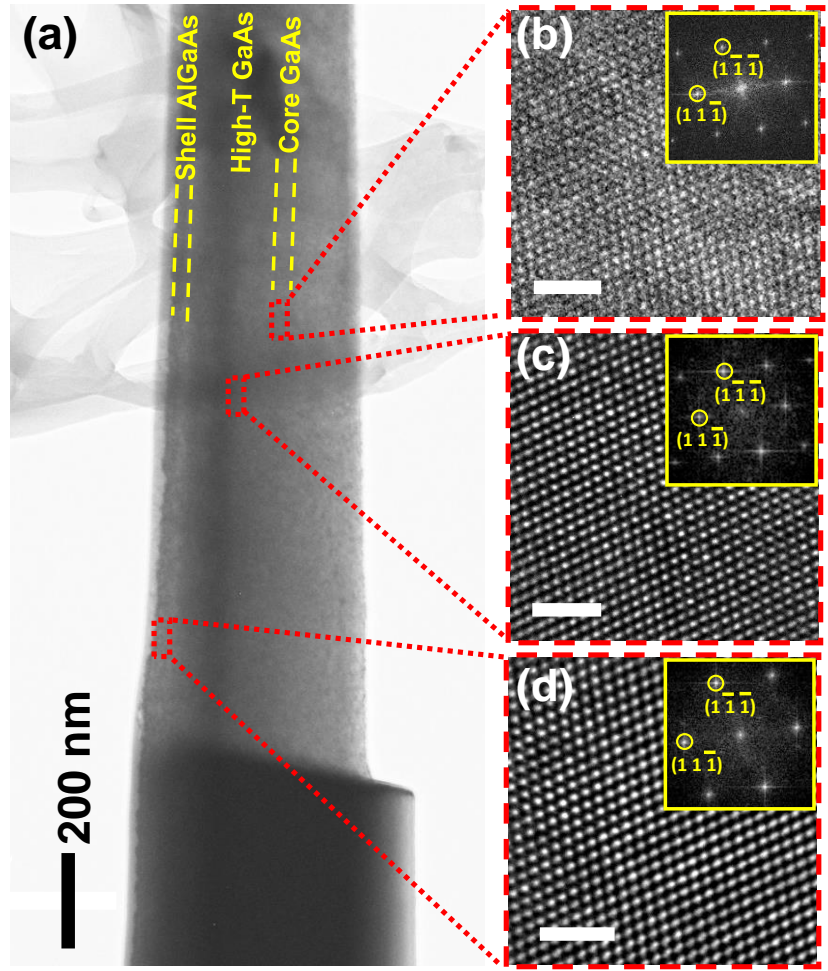

**Figure 2.** (a) TEM image of the core-multishell GaAs/high-T GaAs/AlGaAs nanowire prepared by FIB thinning. The different layers are indicated by yellow dashed lines. (b-d) HR-TEM images taken with <101> viewing orientation show pure ZB structure with excellent crystal quality of all three layers: (b) GaAs core, (c) GaAs inserting shell, (d) AlGaAs shell. Diffraction patterns obtained by FFT are shown as insets. Scale bars are 2 nm.

For optical and transport measurements, nanowires were removed from the growth substrate by ultrasonication in an ethanol bath and dispersed onto a thermal $SiO_2$/Si substrate. Room-temperature micro-photoluminescence spectra[33] of a GaAs nanowire (r=40 nm) and a GaAs/AlGaAs core-multishell nanowire (40/170/30 nm) highlight the role of the AlGaAs shell for surface state passivation and carrier confinement (Fig. 3b). While no obvious emission peak is observed from a single GaAs nanowire (black triangles), the core-shell nanowire structure exhibits a strong photoluminescence peak centered at 878 nm (~1.412 eV, red squares). In this case, the slight red shift of the emission peak compared to bulk ZB GaAs at room temperature (870 nm)[34] may also be an indication of indirect recombination of electrons confined at the GaAs/AlGaAs interface and holes in the GaAs flat valence band region.[35] The corresponding band diagram is illustrated in Fig. 3a.

To elucidate the nature of photoresponse, single MSM nanowire photodetectors were realized from the core-shell GaAs/AlGaAs heterostructure by depositing selective Pt contacts onto the AlGaAs shell and GaAs core using FIB assisted deposition and etching (see Supporting Information for details of the fabrication). An SEM image of the finalized structure with electrodes is shown in Fig. 3c. Representative current-voltage (I-V) measurements are shown in Fig. 3d.[36] Both I-V characteristics (in dark and under illumination) show typical behavior of MSM double Schottky barrier structure, where the asymmetry between positive and negative bias may be due to the different Schottky barrier height between Pt/GaAs and Pt/AlGaAs. The nanowire photodetector displays a photocurrent to dark current ratio of 145 at $V_{SC}$=1 V, which is larger than previously reported GaAs nanowire photodetectors.[24, 37] Under illumination, the photocurrent first increases rapidly in the voltage range of -1 to 1 V, and then continues to grow but less steeply. Under reverse bias, the Schottky barrier height increases and provides a stronger local built-in electric field at the contact interfaces to efficiently separate photo-generated electrons and holes, increasing the photocurrent. The maximum current is limited by carrier drift in the space-charge region near the Schottky contact and at the GaAs/AlGaAs heterointerface for small reverse bias. As the bias increases further, carrier diffusion in the neutral region of the nanowire (i.e. carrier diffusion length and carrier lifetime) then determines the maximum current.[25]

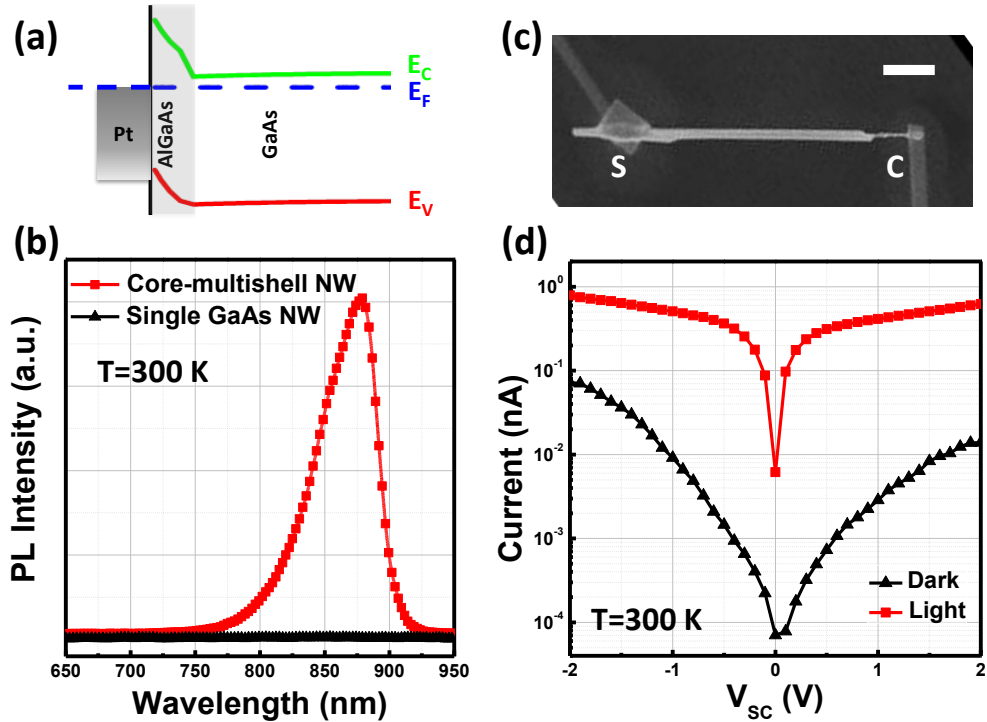

**Figure 3.** (a) Schematic band diagram of the zero-biased metal-semiconductor contact (note that the band discontinuity at the GaAs/AlGaAs interface is masked by the high Schottky barrier induced by the metal contact). (b) Comparison between micro-photoluminescence emission from a single GaAs nanowire and a core-multishell GaAs/high-T GaAs/AlGaAs nanowire at room temperature. (c) SEM image of the MSM nanowire photodetector with selective contacts to the core (C) and to the shell (S). Scale bar is 2 μm. (d) Current-voltage characteristics of the photodetector in the dark and under white light illumination.

The spectral sensitivity of the nanowire photodetector is evaluated by wavelength dependent photocurrent measurements at room temperature.[38] The measured photoresponsivity (Fig. 4a) matches well the typical GaAs spectral photoresponse, with the appearance of few additional photocurrent peaks. This indicates that absorption occurs in the GaAs core rather than in the AlGaAs shell and suggests the existence of resonant optical modes in the nanowire which increase absorption. The nanowire photodetector shows respectable figures of merit: at

$\lambda=855$ nm, the photoresponsivity (the ratio of electrical output to the optical input) is 0.57 A/W, which is even higher than some common GaAs photodetector in the market,[39] and specific detectivity is $7.20\times10^{10}$ cm$\sqrt{Hz}$/W, slightly lower than the one reported for planar GaAs photodetector (details on the derivation of these figures of merit are given in Supporting Information).[40] Spectral shape and amplitude of the experimental photoresponsivity spectrum are in good qualitative agreement with two-dimensional Silvaco-Atlas simulation results obtained by solving the Boltzmann transport equation with charge carrier density and distribution of the 2DET under illumination (Fig. 4b). Moreover, the origin of resonant peaks in the photoresponsivity spectra is unraveled by full-wave optical simulation using COMSOL Multiphysics®. Fig. 4c shows the optical absorption spectrum calculated for our nanowire heterostructure, where at least three peaks can be identified at 655 nm, 800 nm and 850 nm. The strongest of these peaks (mode 1 at 850 nm) can indeed be seen in the experimental spectrum in Fig. 4a, while the higher order modes (mode 2 at 800 nm and 3 at 655 nm) are gradually buried under the noise of the photoconductivity measurements at room-temperature. Optical absorption maps across the section of the nanowire (Fig. 4d-f) help visualizing the spatial distribution of electromagnetic energy within the device, showing how the three resonant modes correspond to optical modes in the GaAs core of the wire, confined by the higher refractive index AlGaAs shell and the air surrounding. This provides further evidence that engineering of the nanowire dielectric properties is an effective tool to increase resonant absorption cross-section and tune spectral sensitivity of nanowire devices.[13, 20, 41-43]

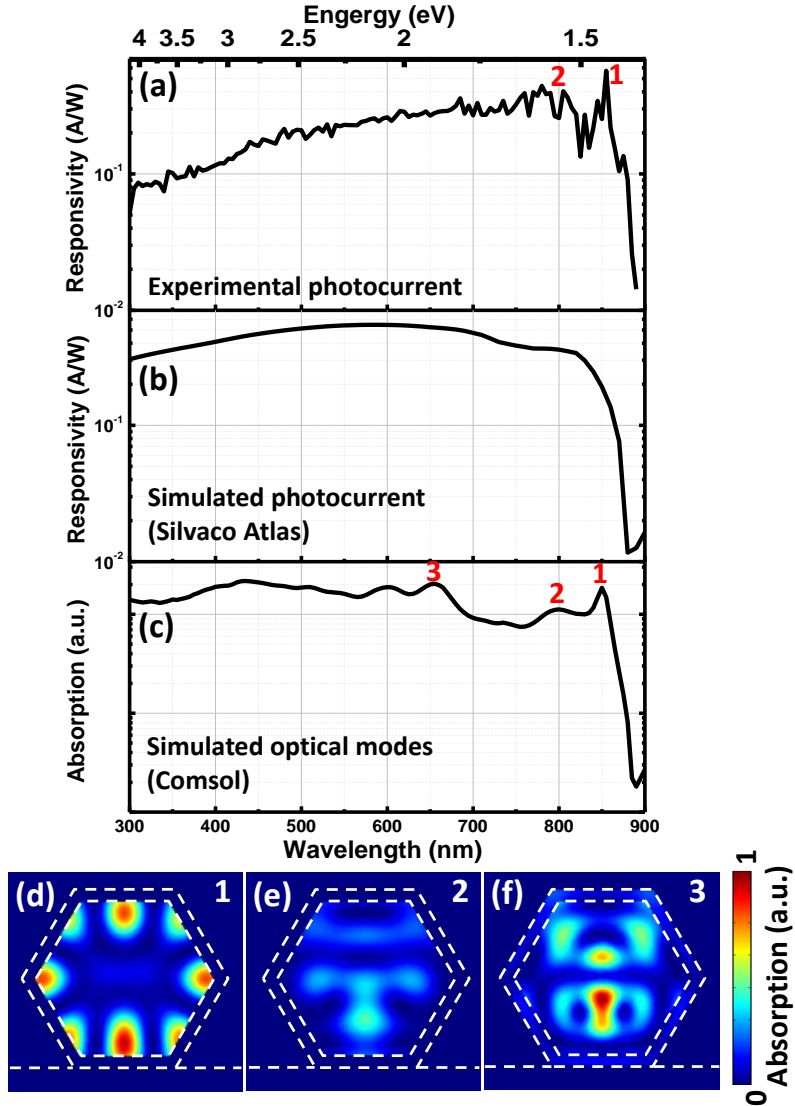

**Figure 4.** (a) Measured spectral photoresponsivity of the core-multishell nanowire photodetector under applied bias of $V_{SC}=2$ V. (b) Simulated photoresponsivity spectrum based on Boltzmann transport model with charge carrier density and distribution of the 2DET under illumination. (c) Simulated optical absorption spectrum based on electromagnetic modeling: resonant absorption peaks, labelled as 1, 2 and 3, emerge at 850, 800 and 655 nm. (d-f) Simulated optical absorption maps across the section of the nanowire at wavelengths corresponding to peaks 1, 2 and 3, respectively.

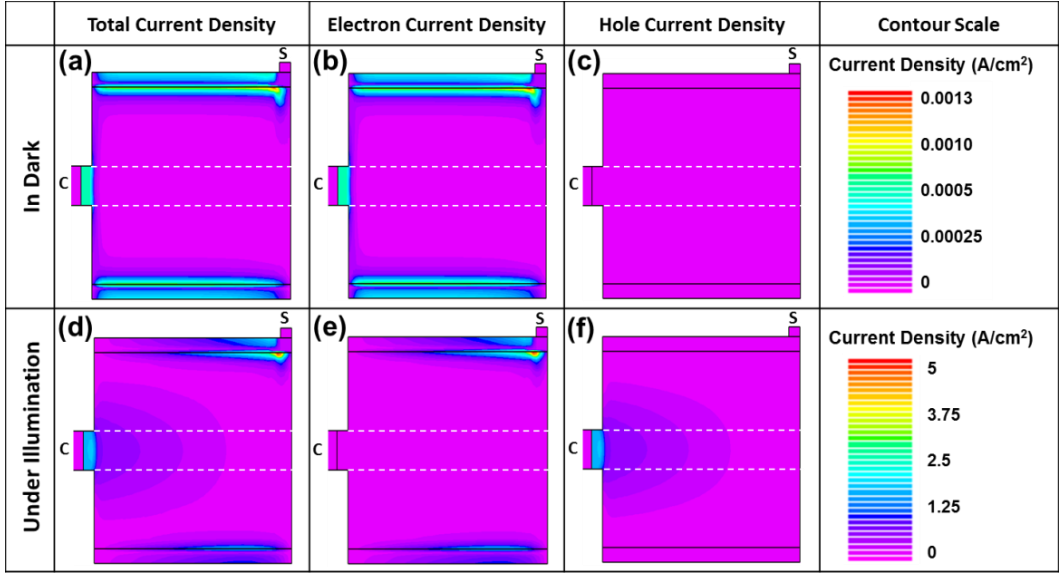

**Figure 5.** Two-dimensional profiles of the current distribution in the core-multishell nanowire photodetector obtained for doping concentrations of $N_D=1\times10^{13}$ cm$^{-3}$ and $N_D=5\times10^{17}$ cm$^{-3}$ in GaAs and AlGaAs, respectively. The simulated core-shell nanowire has diameter of 480 nm and length of 9.5 μm: in the maps scale of the x-axis is compressed by a factor of 22. The top row shows dark current profiles while the bottom row shows the photocurrent distribution under illumination. (a) and (d) are total current density (electron plus hole current), (b) and (e) electron current density, and (c) and (f) hole current density plots. Dashed lines indicate the nanowire core region.

Finally, to understand the spatial distribution of photogenerated carriers within the nanowire heterostructure, hole and electron densities were calculated using Silvaco-Atlas simulation tool in the dark and under typical illumination conditions used in the experiments. Fig. 5 shows the total (electron plus hole), electron and hole current densities computed at $V_{SC}=2$ V. In the dark, the total current density is low and mainly contributed by the electron current (Fig. 5a-c). The core-to-shell contact configuration suppresses the 2DET at the bare core end, reducing the dark current. Under illumination, photons are primarily absorbed in GaAs, producing electron-hole pairs. Photogenerated electrons are accumulated at the GaAs/AlGaAs interface (Fig. 5e), while holes are stored at the boundary between the electrode and GaAs due to the raised Schottky barrier (Fig. 5f). Holes are then collected by the core electrode while electrons are transferred along the 2DET channel and collected by the shell electrode. Spatial separation of electron and holes induced by the core-multishell nanowire structure with MSM electrode configuration thus reduces carrier recombination and increases the total photocurrent (Fig 5d), validating our design concept for efficient NW photodetectors.

In conclusion, a novel core-multishell GaAs/high-T GaAs/AlGaAs nanowire photodetector structure with MSM electrode configuration is proposed and demonstrated. The realized structure exhibits excellent crystallographic properties with pure ZB phase and low defect density. Surface states are effectively passivated by the AlGaAs shell, as evidenced by the enhancement of radiative emission from a single nanowire at room temperature. Thanks to the core-to-shell contact geometry, the device yields low dark current and high photoresponsivity in the wavelength range of 300 to 890 nm, with large photocurrent to dark current ratio. Self-consistent solution of Schrödinger-Poisson equations demonstrates the presence of a 2DET at the GaAs/AlGaAs interface, while carrier transport simulations reveal that electrons are collected at the core-shell interface while holes are accumulated at the Schottky contact between electrode and GaAs core, reducing carrier recombination. Full wave electromagnetic modeling shows the presence of resonant optical modes supported by the waveguiding properties of the dielectric wire, which selectively enhance photocurrent at the resonant wavelengths. Overall this study highlights the potential of combined optimization of crystallographic, optical and transport properties to design highly efficient III-V nanowire photodetectors.

**Acknowledgements**

The work was supported by Nanyang Technological University (project reference M4080538 and M4080511) and by the Singapore Ministry of Education (project reference MOE2013-T2-

044 and MOE2011-T3-1-005). The authors are grateful to Prof. Tang Xiaohong for his assistance with MOCVD operation. C.C. would like to thank the Champagne-Ardenne region for financial support via the 'visiting professor' scheme and the Emergence project 'NanoGain'.

following: R(λ)=$I_{ph}$/P where $I_{ph}$ is measured photocurrent and P is the light intensity of each wavelength.

**TOC graphic**

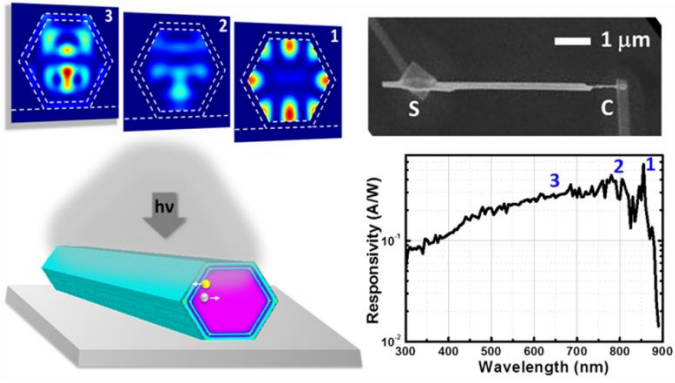